\documentclass[traditabstract]{aa}
\usepackage{graphicx}
\usepackage{subfloat}
\usepackage{graphicx}
\usepackage{caption}
\usepackage{subcaption}
\usepackage{amsmath}
\usepackage{natbib}
\usepackage{color}
\usepackage[english]{babel}
\bibpunct{(}{)}{;}{a}{}{,} 

\begin{document}
\title{Hiding its age: the case for a younger bulge}
\titlerunning{Hiding its age: the case for a younger bulge}

\author{M. Haywood\inst{1}, P. Di Matteo\inst{1}, O. Snaith\inst{2}, A. Calamida\inst{3}}

\authorrunning{Haywood et al.}

\institute{GEPI, Observatoire de Paris, CNRS, Universit$\rm \acute{e}$  Paris Diderot, 5 place Jules Janssen, 92190 Meudon, France\\
\email{Misha.Haywood@obspm.fr}
\and 
School of Physics, Korea Institute for Advanced Study, 85 Hoegiro, 
Dongdaemun-gu, Seoul 02455, Republic of Korea  
\and
National Optical Astronomy Observatory/AURA, Tucson, AZ 85719
}

\date{Accepted, Received}

\abstract{
The determination of the age of the bulge has led to two contradictory results. 
On the one side, the color-magnitude diagrams in different bulge fields seem to indicate 
a uniformly old ($>$10~Gyr) population. On the other side, individual ages derived from dwarfs observed through 
microlensing events seem to indicate a large spread, from $\sim$ 2 to $\sim$ 13 Gyr.
Because the bulge is now recognised as being mainly a boxy peanut-shaped bar, it is suggested that disk stars are 
one of its main constituents, and therefore also stars with ages significantly 
younger than 10 Gyr. Other arguments as well point to the fact that the bulge cannot be
exclusively old, and in particular cannot be a burst population, as it is usually expected 
if the bulge was the fossil remnant of a merger phase in the early Galaxy. 
In the present study, we show that given the range of metallicities observed in the bulge, 
a uniformly old population would be reflected into a significant spread in color at the turn-off which is not 
observed.
Inversely, we demonstrate that the correlation between age and 
metallicity expected to hold for the inner disk would conspire to form a color-magnitude diagram with a remarkably
small spread in color, thus mimicking the color-magnitude diagram of a uniformly old population. If stars younger than 10 Gyr are part 
of the bulge, as must be the case if the bulge has been mainly formed through dynamical instabilities
in the disk, then a very small spread at the turn-off is expected, as seen in the
observations. 
}

\maketitle

\section{Introduction}
Progress in the study of galactic bulges in the last decade have established that they come
in two different flavours: classical and pseudo-bulges \citep{kor04}. 
Classical bulges have properties identical to elliptical galaxies, while pseudo-bulges
have properties inherited from the disk from which they are supposedly formed: they are more flattened, have bluer colors, and rotate faster. 
In the local volume (distance $<$ 11Mpc), pseudo-bulges are the majority, 
and classical bulges are few \citep{kor10,fis10}, while some galaxies are known
to harbour the two types of bulges \citep{erw15, fis16}.
It has been found first from kinematic
constraints that the classical bulge must be small or inexistent \citep{she10,kun12,dim14} 
in the  Milky Way (hereafter MW).
At the same time, arguments based on kinematic and chemical properties of stellar populations 
observed in the bulge of the MW have strengthen the 
picture of a pseudo-bulge with clear correspondence between the stellar populations
identified in the bulge and those of the inner disk of the Galaxy. 
Hence, in the nomenclature of \cite{nes13}, it has been suggested that components A, B and C could be 
identified to the thin disk, young and old thick disk of the MW (Di Matteo et al. 2014, Di Matteo 2016).  
The expectations, based on the recognition that the bulge of the 
MW is a pseudo-bulge, together with the 
chemical arguments, are therefore that most of the bulge have a disk origin, implying that it should contains 
stars of all ages \citep{nes13}.

\cite{cla08} (hereafter CL08) selected a pure sample of bulge stars in the Sagittarius Window Eclipsing Extrasolar Planet Search (SWEEPS) field by using proper motion measurements based on Hubble Space Telescope data.
They obtained the color-magnitude diagram (CMD) which is to date the strongest 
support for an old bulge, showing a tight old turn-off and no significant upper
main sequence. The very tight CMD in the region of the turn-off has been confirmed
in \citeauthor[(2015, hereafter CA15)]{cal15}.
\cite{ben11,ben13} have measured metallicities and 
abundances of microlensed  dwarfs stars in the bulge, estimating their age. They found
that the bulge shows a large spread in age, with stellar ages ranging from less than 3 up to 13 Gyr. 
While \cite{cla11} estimate that not more than 3.5\% of bulge stars may have 
an age smaller than 5 Gyr, this percentage rises to 22\% for the microlensed dwarfs studied by \citeauthor[(2013, hereafter B13)]{ben13}.
Clearly, stars younger than about 10 Gyr are a significant fraction of the bulge according to
the results of B13.
The present picture is therefore contradictory, and the arguments seem to be incompatible 
\citep[see also][]{nat15}.

In their study, B13 suggest that a correlation between age and metallicity 
could conspire to produce stars which would be located in a relatively narrow 
sequence in the HR diagram, giving the impression of a co-eval, old population, as observed
by CL08. 
This idea has received little attention however, possibly because one may think that such 'conspiracy'
would require a very specific, and ad-hoc bulge chemical evolution. 
In the present study, we first show that an exclusively old bulge, combined with the 
large spread in metallicity that is observed in the bulge, cannot 
produce the tight turn-off observed in the HST CMD. 
We then show that if the metal-rich stars are younger, the disagreement is considerably 
reduced, and obtaining a dispersion compatible with the one observed is becoming possible. 
The problem is not that the apparently old bulge turn-off 
is incompatible with significantly young stars, but to understand how and how much
young stars can hide in the bulge CMD.

The outline of the present article is as follows. We first introduce our chemical evolution model 
of the inner disk and the age-metallicity relation (hereafter AMR) obtained from this model. A detailed comparison with 
the observed properties of the inner disk and the bulge can be found in Haywood et al. (in preparation).
We then give a brief description of the
properties of the SWEEPS's Window (CL08) and the constraints from the observed CMDs. 
In Section  3 we explain how we reconstruct the CMD of the bulge 
when assuming the AMR of our model or a co-eval population. 
In the third section, we present realistic simulations of the CMD of the bulge which are 
discussed in the light of the known characteristics of the observed CMDs. 
In the last section, we discuss our results in the context of the bulge evolution 
and review the other arguments that 
have been invoked to claim that the bulge is exclusively old.

\begin{figure}
\includegraphics[width=9.cm]{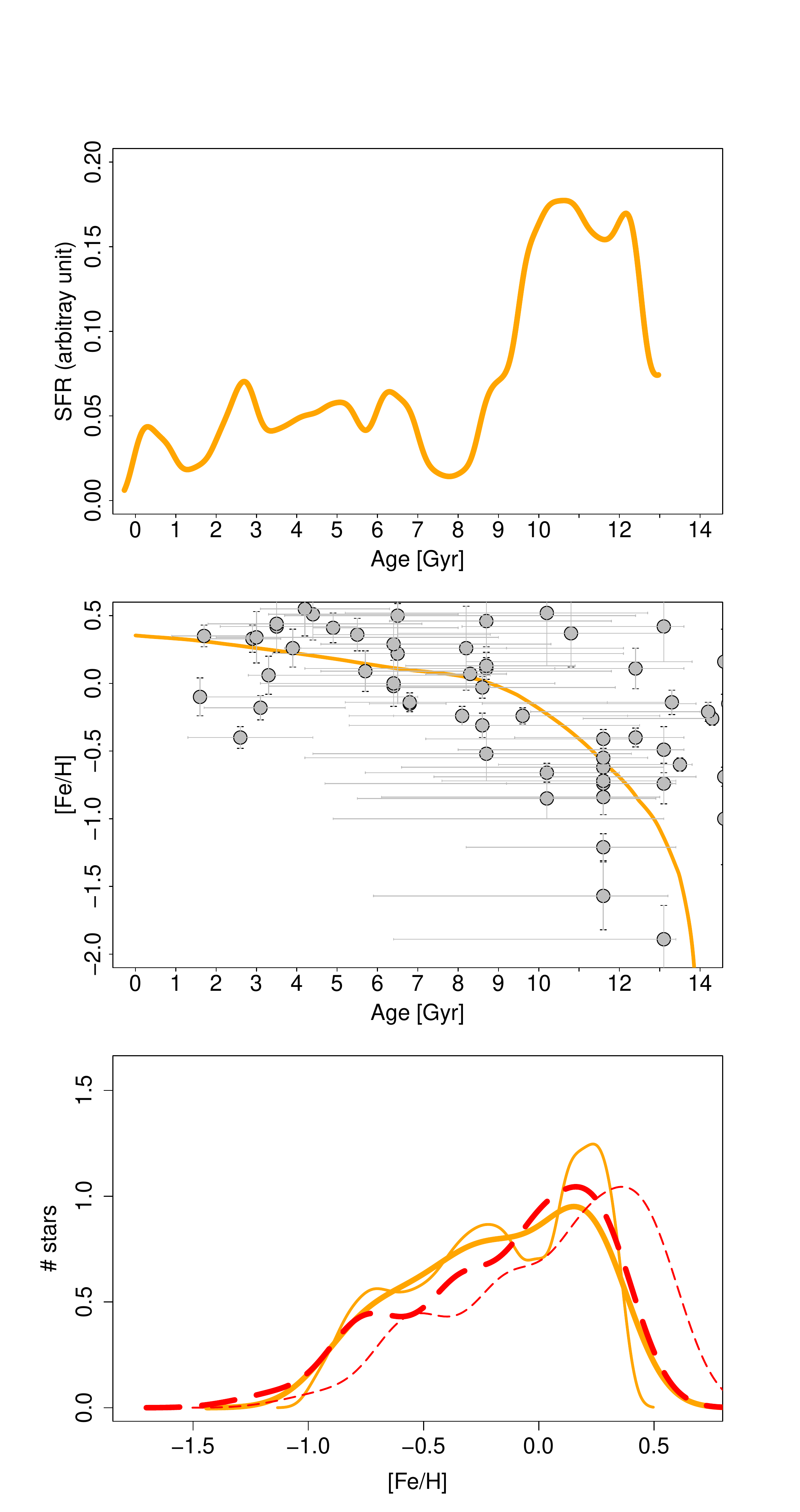}
\caption{Top: Star formation rate history of the model. Middle: age-metallicity relation of the model, together with 
the stellar age and metallicities of the microlensing  events of bulge stars from B13.  
Bottom: the metallicity distribution 
from the model (thin orange curve) and convolved with a gaussian of 0.15 dex dispersion (thick orange curve) 
together with the generalised histograms of the observed MDF in the Baade's Window 
from Hill et al. (2011) (thin dashed red curve), and shifted by -0.21 dex (thick dashed red curve),
following the results of \cite{roj14}.
}
\label{fig:model}
\end{figure}

\section{The chemical evolution model and observational constraints}

We aim at understanding under which conditions on the age distribution of stars the CMD of the bulge, 
and particularly the CMD obtained by CL08 or CA15 can be reproduced. 
To first order, the density of stars in a color-magnitude diagram results from 
its star formation history (masses and ages) and correlated chemical evolution (metallicities and abundances). 
This can standardly be represented by an initial mass function (IMF), a star formation history (SFH), 
and the variation of metallicity and element 
abundances with age. We therefore need to adopt a model which describes these evolutions.

\begin{figure}
\includegraphics[width=9.cm]{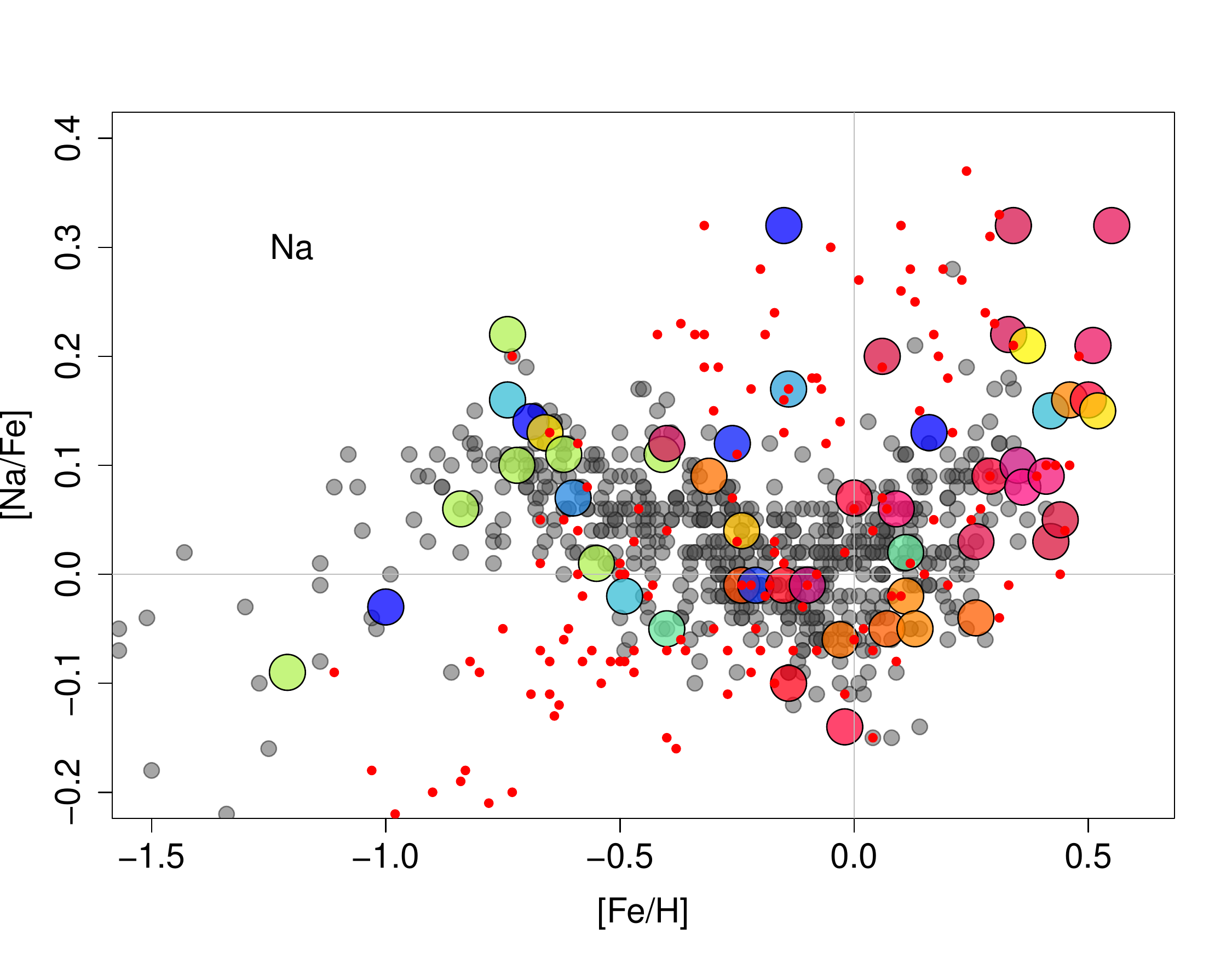}
\caption{
Sodium abundance ratio as a function of [Fe/H] for stars in the disk (gray dots)  from \cite{ben14} 
and for stars in the bulge (larger symbols) from B13. 
The data from \cite{joh14} are shown as red dots. 
}
\label{fig:nafeh}
\end{figure}

\subsection{Chemical evolution model}

In \cite{sna15} we described a model which was fitted to the thick and thin disk data 
of stars at the solar vicinity.
A SFH of the inner disks was derived by fitting the model to the age-[Si/Fe] relation found in \cite{hay13} 
on solar vicinity stars.
In \cite{hay15}, we showed that this age-[Si/Fe] relation is a general property  of the 
inner disk, implying that the SFH of \cite{sna15} is also valid for the whole inner disk.
In \cite{hay16}, we showed further that our model provides a very good
match to the inner disk APOGEE abundance data.
Improvements in the observations of the bulge in the last ten years have shown that the alpha elements, while originally 
believed to be significantly higher than those of the solar vicinity \citep[see e.g.][]{lec07} are now
measured to be near \citep{gon15}, or similar (B13) to solar abundances, adding strong
support to the idea that chemical evolution models could fit the bulge and disk chemical properties at the same time.
In an accompanying study (Haywood et al. in preparation), we show indeed that the chemical properties of the bulge 
(chemical abundance patterns and metallicity distribution) are compatible with those 
to the inner thin disk and the thick disk of the MW, and that these properties can be well described by the same 
model which we used to represent the chemical evolution of inner disk stars.
In a scenario where most of the bulge forms by dynamical instabilities in the disk and its stars originate 
in the thick disk and inner thin disk, this is expected, and justifies that 
we use the age-metallicity relation and the SFH of our model as input parameters in building the 
CMDs of the bulge.

The ingredients of the model are described and discussed in detail in \cite{sna15}.
These ingredients are: standard theoretical yields \citep{iwa99,nom06,kar10}, an IMF
\citep{kro01} independent of time, and a stellar mass--lifetime relation dependent on the metallicity \citep{rai96}.
The \cite{kro01}  IMF is in good agreement with the IMF estimated for the bulge in CA15.
No instantaneous recycling approximation has
been implemented, see \cite{sna15} for details.

Fig.\ref{fig:model} shows the main distributions of the model resulting from fitting the age-[Si/Fe] relation: the star formation history (a), 
age-metallicity relation (b)
and metallicity distribution function (c).
Fig. \ref{fig:model}(b) compares the age-metallicity relation of the model with the data on microlensing events 
from B13. 
The MDF of the model (thin and thick continuous curves) is compared to the observed MDF from Hill et al. (2011, dashed thin curve, Fig. \ref{fig:model}(c)) in the Baade's Window.
A systematic shift is apparent between the model and observed MDFs, of about 0.2 dex. 
More recently, \cite{roj14}  compared the metallicities of about 100 stars they have in 
common with \cite{hil11}, and find an offset of 0.21 dex. 
The thick dashed red curve shows the \cite{hil11} MDF corrected by this amount. The observed MDF now shows good
agreement with the model. The thick continuous curve shows the model convolved with a gaussian of 0.15 dex dispersion
in order to take into account the measurement errors of observed metallicities.
A detailed discussion on the comparison between abundances of the bulge and disk stars is beyond the present paper
and is provided in \cite{hay16}. We comment however on the particular 
abundance of sodium, which is presented as the best evidence of a different chemical histories of the 
bulge and disk, on the basis of the study of \cite{joh14}.
Fig. \ref{fig:nafeh} shows the abundance data of dwarfs in the solar vicinity from Bensby et al. (2014, grey dots), 
microlensed dwarfs of B13 and giants from \cite{joh14}. 
The figure illustrates that while dwarf data of the bulge and solar vicinity show good agreement, 
the bulge giants of \cite{joh14} and bulge dwarfs of B13 (red and gray
dots respectively) have very different distributions. Because the data of B13 and \cite{ben14}
are both obtained on dwarfs (and measured from the same authors), they are less prone to possible systematics.
This plot illustrates that Na abundances of the bulge can be affected by significant uncertainties and the 
difference seen by \cite{joh14} by comparing giants of the bulge and disk dwarfs of the solar 
vicinity may only be the effect of residual either between giants and dwarfs or in reduction 
procedures. We therefore conclude that sodium abundances of \cite{joh14} provides no conclusive evidence 
of a real difference between the disk and bulge nucleosynthesis history, while the other two datasets 
of the bulge and disk dwarfs show excellent agreement.

The SFH, age-metallicity relation and MDF distributions of Fig. \ref{fig:model} are used to generate the CMDs discussed in Sec. 3.

\subsection{The Galactic  bulge SWEEPS field}

In the bulge, spatial segregation occurs, with the most metal-poor (presumably older) populations 
dominating at larger distances from the Galactic plane, while in-plane fields are dominated
by the most-metal-rich (presumably the youngest) component \citep{nes13}. The field observed
by CL08, \cite{cla11} and CA15 at l=1.25$^{\circ}$ and b=-2.65$^{\circ}$ 
is 340pc below the Galactic plane at the distance of the bulge (8kpc). Hence, although the field is 
above the star formation 
layer, it  may contain a significant number of stars a few Gyr old. 
However, young stars, if present, should be visible above the turn-off 
in the SWEEPS field, but no significant number have been detected (see CL08).
Therefore, in the case where we consider that the bulge may have a significant spread in age, we assume that stars are  
3 Gyr old or older.

In comparing with the data, a key ingredient is the metallicity distribution function. 
The MDF of the bulge is known to be a strong function of Galactic latitude \citep[e.g.][]{nes13}, 
with the metal-rich component contribution decreasing at increasing latitude, but the 
spread in metallicity is large anywhere \citep{nes13,gon15}.

According to \cite{nes13}, the contribution of the metal-rich component (roughly [Fe/H]$>$0.0 dex, 
their component A) is already as important as the medium metal-poor component \citep[component B in][]{nes13} at b=-5$^{\circ}$. 
In Baade's Window, the contribution of the metal-rich component is dominant (see e.g, Gonzalez et al. 2015),
with a strong peak at [Fe/H]$\sim$+0.25 dex, very similar to the model distribution of Fig. \ref{fig:model}, and a spread from
[Fe/H]=-1 to 0.7 dex.  
Metallicities measured from medium-resolution spectroscopy in the SWEEPS field 
for 93 stars show a similar spread from -1 to 0.8 dex, 
also compatible with other previous studies (\citealp{hil11}; B13; \citealp{nes13}, see Calamida et al. 2014 for more details).
Our model MDF reproduces fairly well these main features: a large spread in metallicity, and 
a prominent peak at super-solar metallicities, or [Fe/H]$\sim$+0.20 dex. 
Metallicities below -1.0 dex represent 5\% of the total in the model MDF, while observed MDF
often have a negligible amount of stars below this limit. For a part, this may be due 
to the small statistics of most surveys (a few hundred stars). In \cite{nes13}, where 
the number of objects is more than one order of magnitude higher 
than in other surveys, the tail at [Fe/H]$<$-1.0 dex is clearly visible, weighting 
about 1$\pm$1\% at b=-5$^{\circ}$ \citep{nes13}.
In the following, we adopt the MDF of our model as a realistic representation  
of the MDF of the bulge in the Baade's Window or nearby SWEEPS fields, but also evaluated the 
photometric dispersion at the turn-off when discarding stars with [Fe/H]$<$-1.0 dex.

\begin{figure}
\includegraphics[width=9.cm]{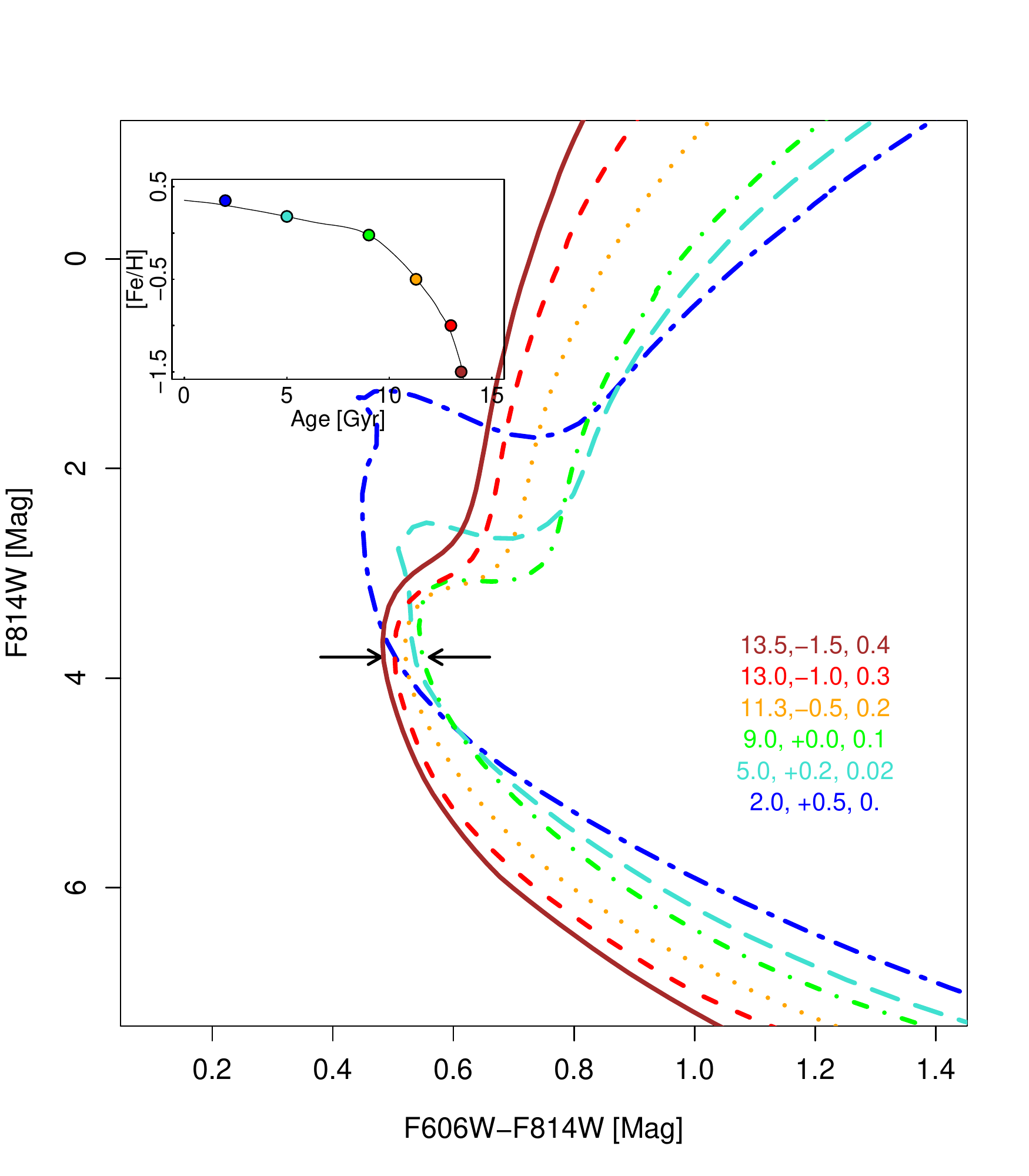}
\caption{Case (a): isochrones along the AMR of Fig. \ref{fig:model}, also inset for ages 5, 7, 9, 10 and 13 Gyr.
Corresponding metallicities and alpha abundances are given on the plot. The two arrows at the turn-off
are separated by 0.07 mag. The age, metallicity and [$\alpha$/Fe] abundance are given for each isochrone. 
}
\label{fig:dhramr}
\end{figure}

\begin{figure}
\includegraphics[width=9.cm]{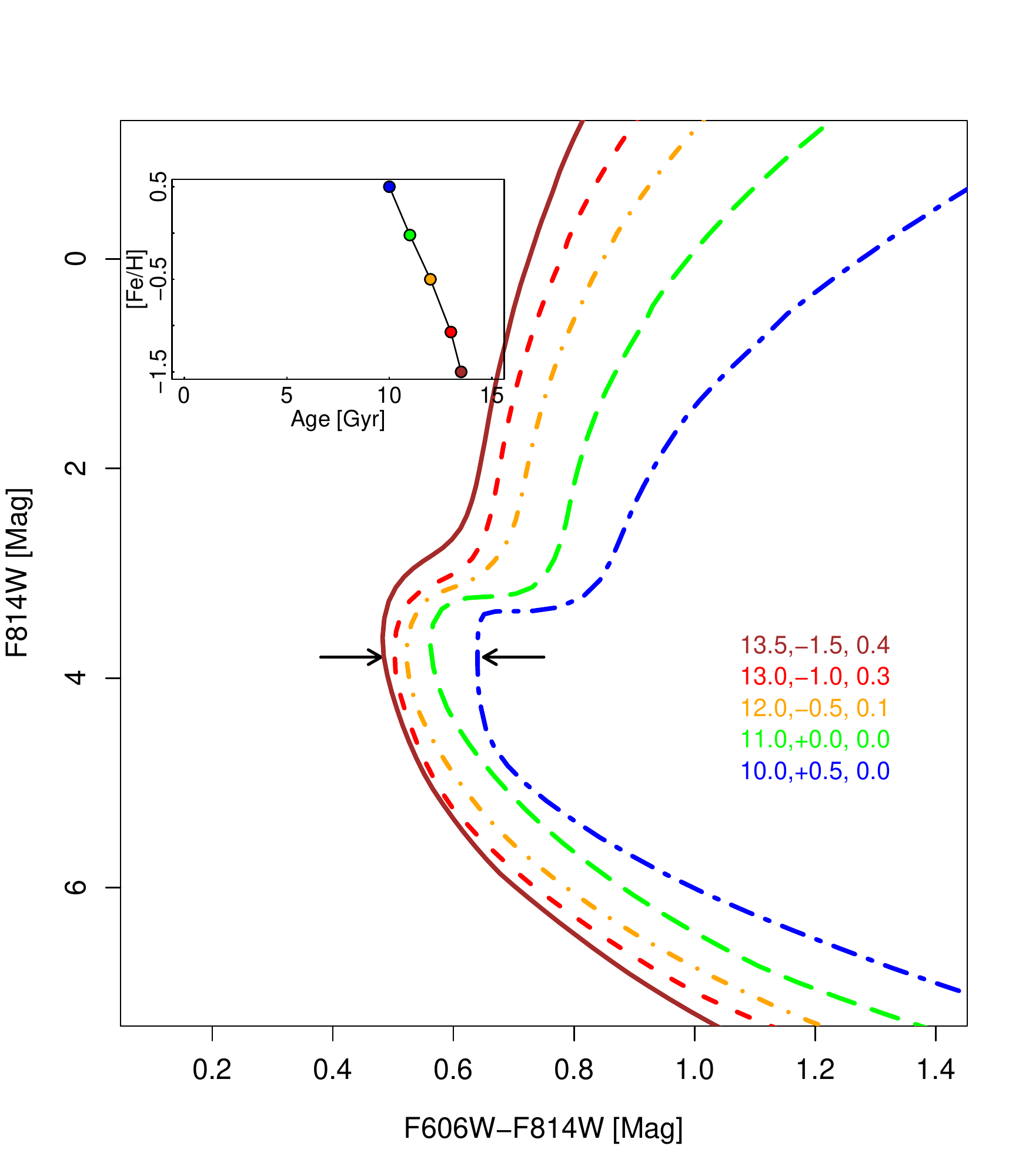}
\caption{Case (b): isochrones at 10, 11, 12, 13 and 13.5 Gyr and metallicities +0.3,0.0, -0.5, -1.0 and -1.5 dex. 
The level of alpha elements is given on the plot. The two arrows at the turn-off
are separated by 0.17 mag. The age, metallicity and [$\alpha$/Fe] abundance are given for each isochrone. 
}
\label{fig:dhrmet}
\end{figure}

 \begin{figure}
\includegraphics[width=9.cm]{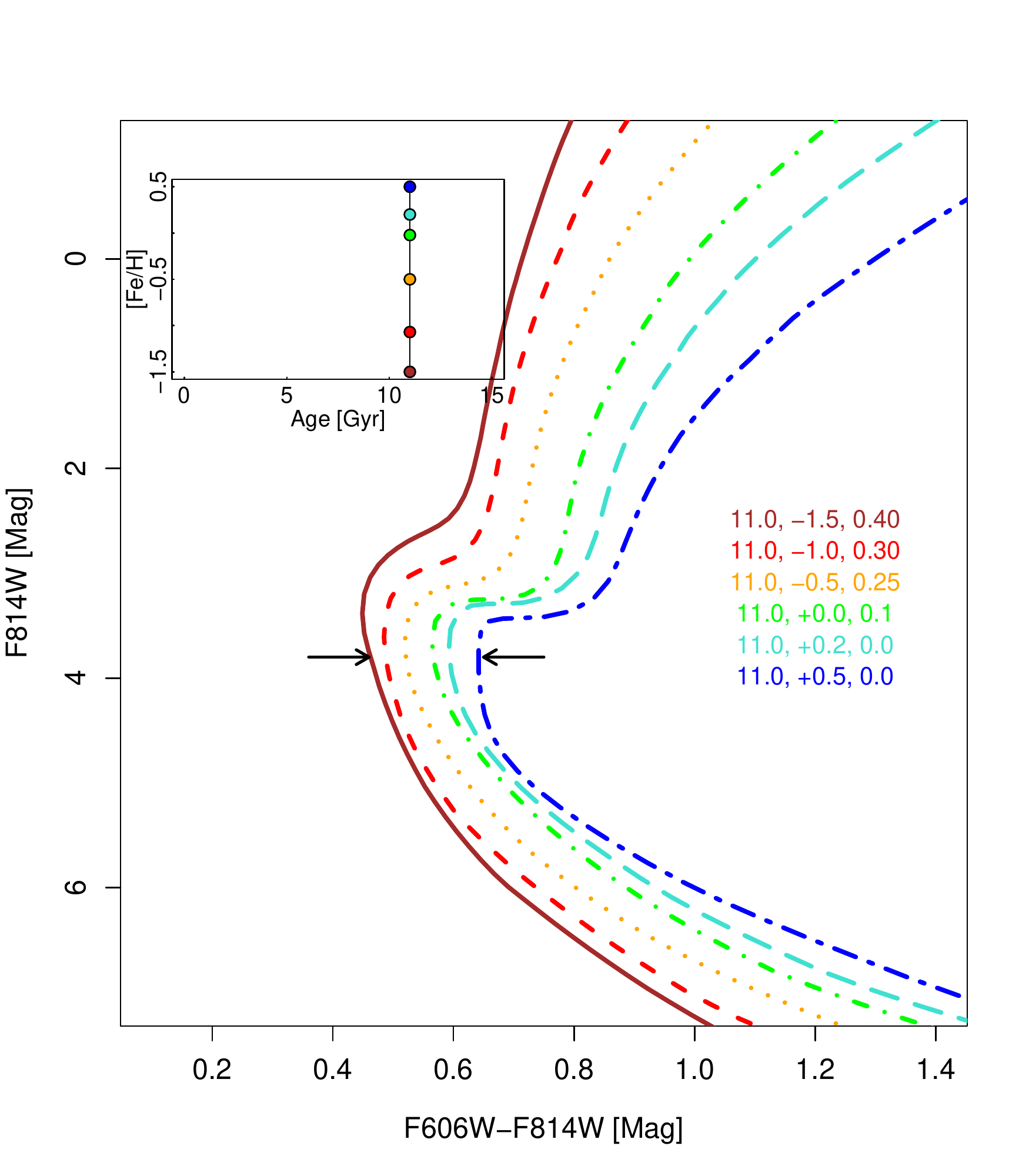}
\caption{Case (c): isochrones at age 11 Gyr and metallicities ranging from 
-1.5 to +0.5 dex. The two arrows at the turn-off
are separated by 0.19 mag. The age, metallicity and [$\alpha$/Fe] abundance are given for each isochrone. 
}
\label{fig:dhriso}
\end{figure}

In the CMD of CA15 (Fig. 4), the dispersion at the turn-off is 0.031 mag, and the spread, 
which we take as three standard deviations, is 0.093 mag\footnote{In the remaining of the article, 
the term 'dispersion' is used for standard deviation, while the 'spread' is assumed as being 3 times the 
dispersion, and is taken as representative of the observed spread at the turn-off.}. 
The contribution of the photometric and data deduction technique to the spread is estimated to be 0.028 mag. 
Deducing this contribution of the observed dispersion, we obtain a dispersion of 0.0295 mag. 
This dispersion is due to age, metallicity, contribution of binaries, differential extinction, and spread in
distances of the stars. 
The intrinsic (due only to age and metallicity effects) dispersion at the turn-off is therefore extremely small
and lower than 0.03 mag. We now seek to understand at what condition on age a synthetic CMD is compatible with such 
a narrow turn-off given the very wide distribution of metallicities observed in the bulge, and taking into account
all the effects just mentioned through Monte-Carlo simulations.

\begin{figure}
\includegraphics[trim=40 140 10 110,clip,width=9.cm]{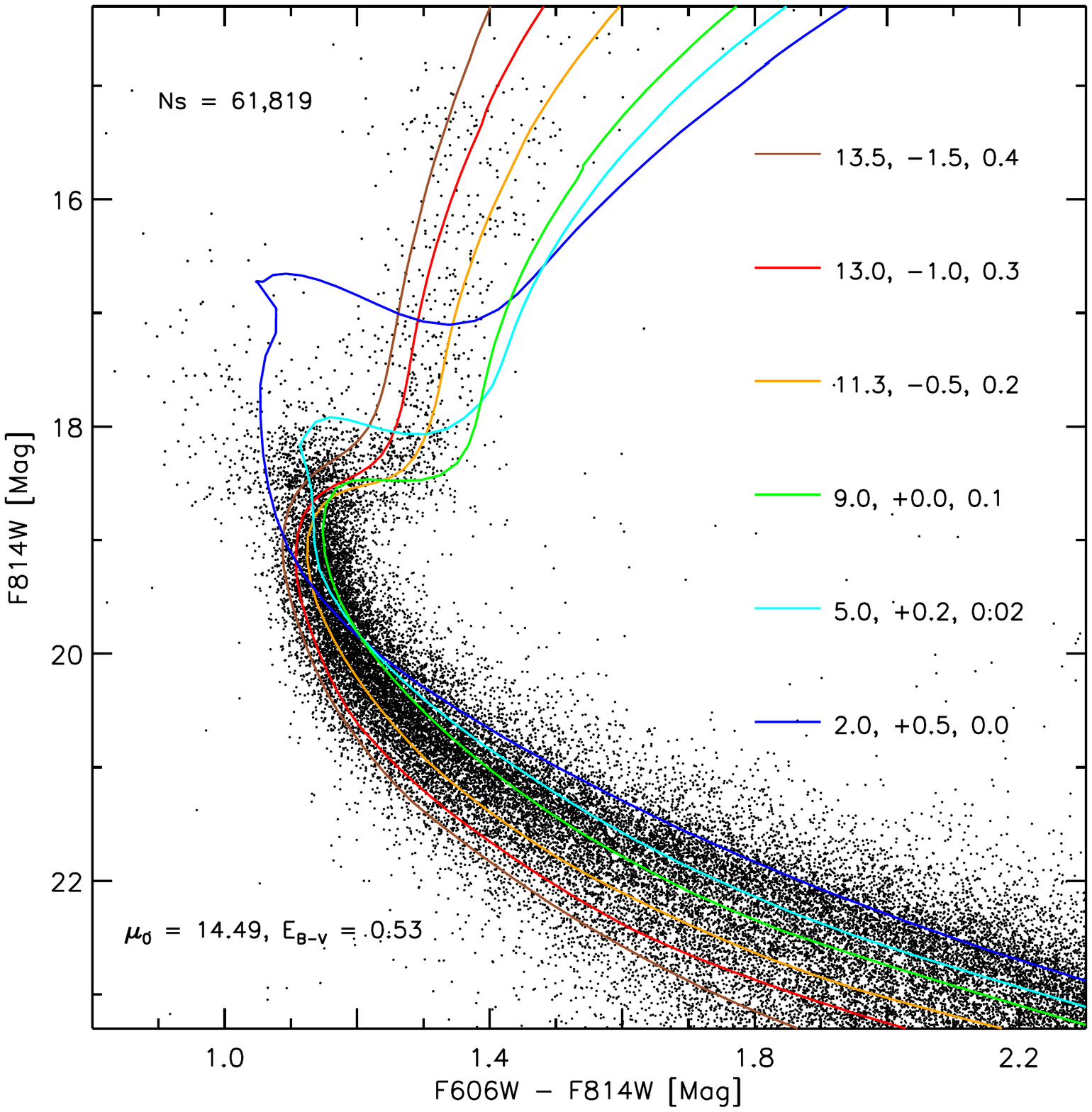}
\caption{
CMD in the SWEEPS field together with the isochrones of Fig. \ref{fig:dhramr}.
}
\label{fig:cmd_age}
\end{figure}

 \begin{figure*}
\includegraphics[width=16.cm]{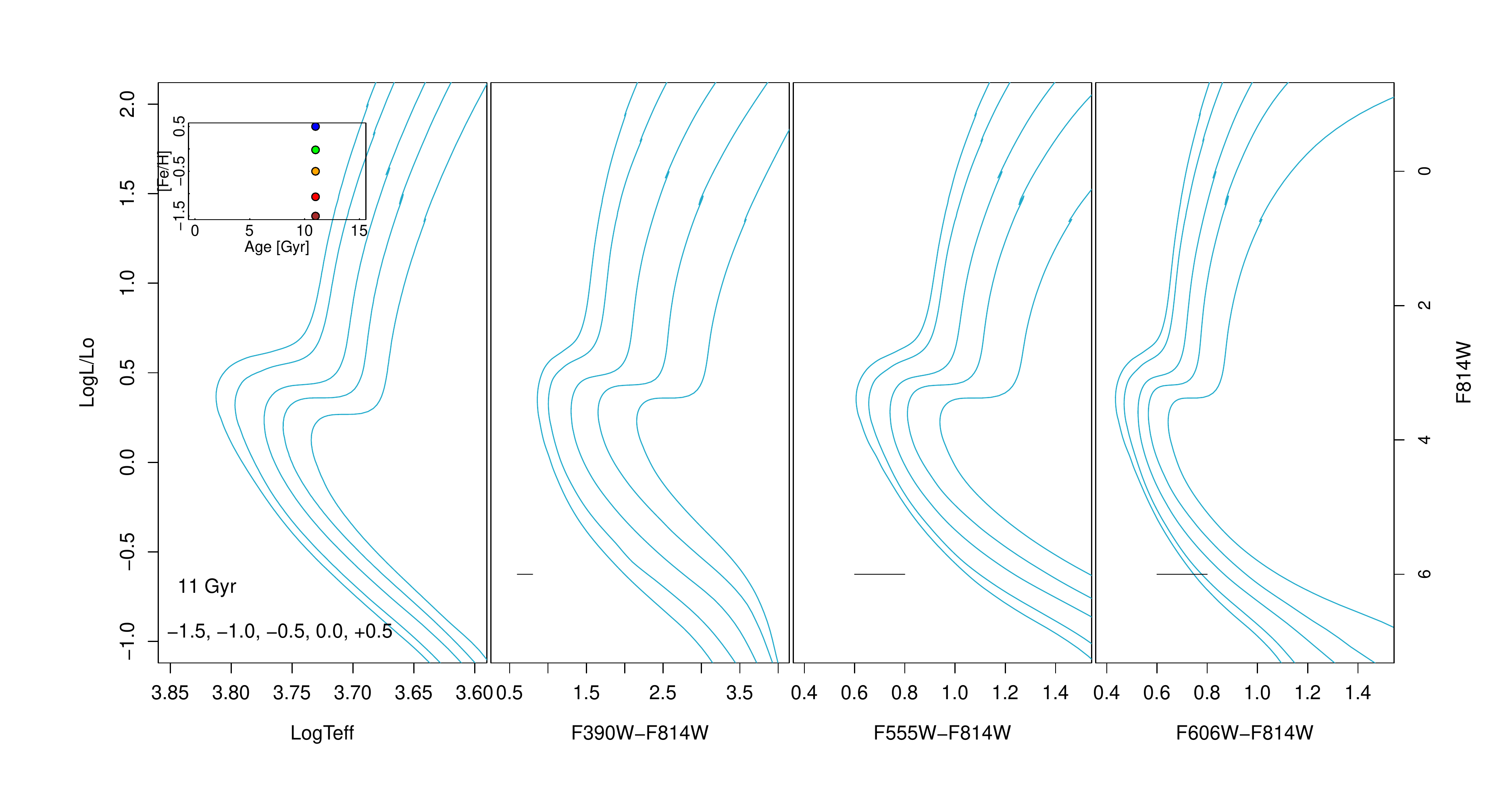}
\includegraphics[width=16.cm]{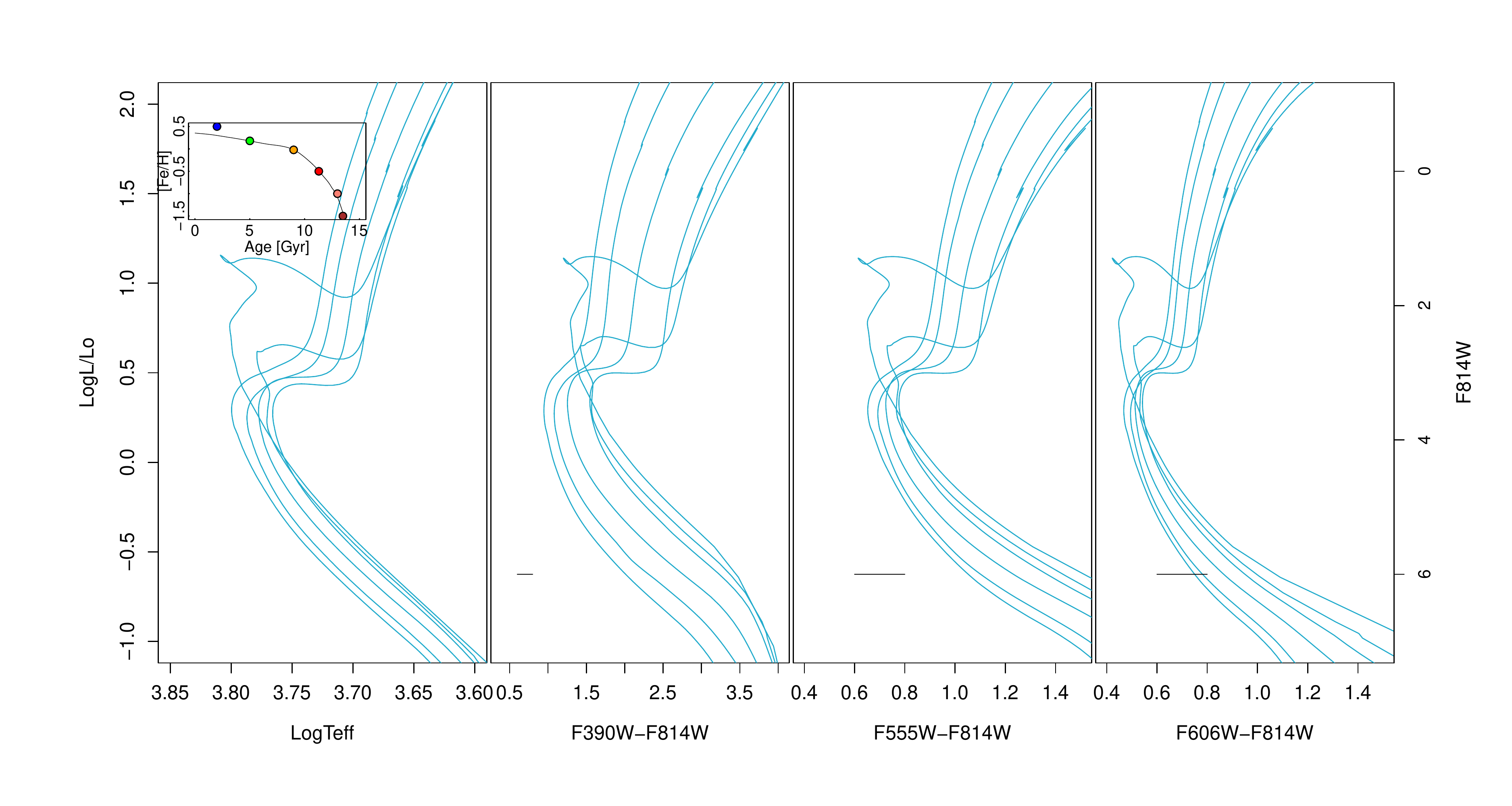}
\caption{Isochrones from Dartmouth at age 11 Gyr and metallicities ranging from 
-1.5 to +0.5 dex (top) and along the age-metallicity relation of Fig. \ref{fig:dhramr} (bottom), 
in the theoretical CMD in different WFC3 colors. The horizontal short 
segment in each CMD is 0.2 mag long, to compare with the width of the turn-off
in each color band. The color separation at turn-off between the most metal-rich and metal-poor
isochrones is 0.21 mag (top) and 0.074 mag (bottom) in the F606W-F814W, similarly to Fig. \ref{fig:dhriso} and \ref{fig:dhramr}. 
The color separation at the turn-off is 0.333 and 0.124 mag in the F555W-F814W bands. 
Alpha abundances of the isochrones are the same as those used for the isochrones of Fig. \ref{fig:dhriso}
and Fig. \ref{fig:dhramr}.
}
\label{fig:dartiso}
\end{figure*}

\begin{figure}
\includegraphics[trim=0 60 10 110,clip,width=9cm]{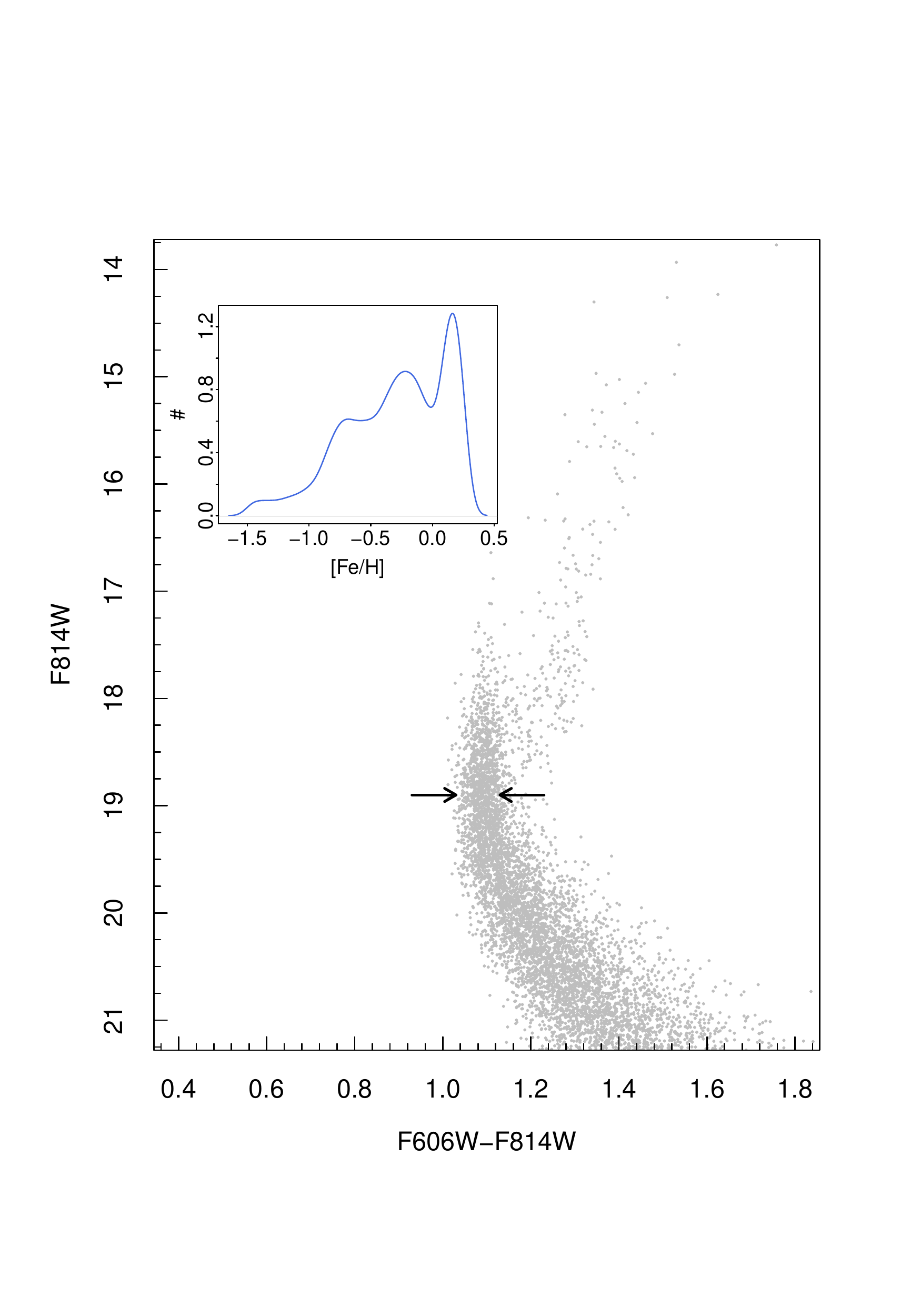}
\caption{Monte Carlo simulations  of the CMD
of Fig. \ref{fig:dhramr} in the F814W/F606W-F814W bands, as described in the text. 
A dispersion of 0.26 mag has been applied on magnitudes to take into account the spread in distance,
0.016 mag on colors to take the differential extinction into account, and 30\% binaries.
The two arrows on each plot correspond to separation of 0.10 mag. The inset frame shows
the corresponding metallicity distribution of the simulation. The percentage of objects younger than 
9 gyr in this plot is 34\%.
}
\label{fig:hstdhr_amr}
\end{figure}

\begin{figure}
\includegraphics[trim=0 60 10 110,clip,width=9cm]{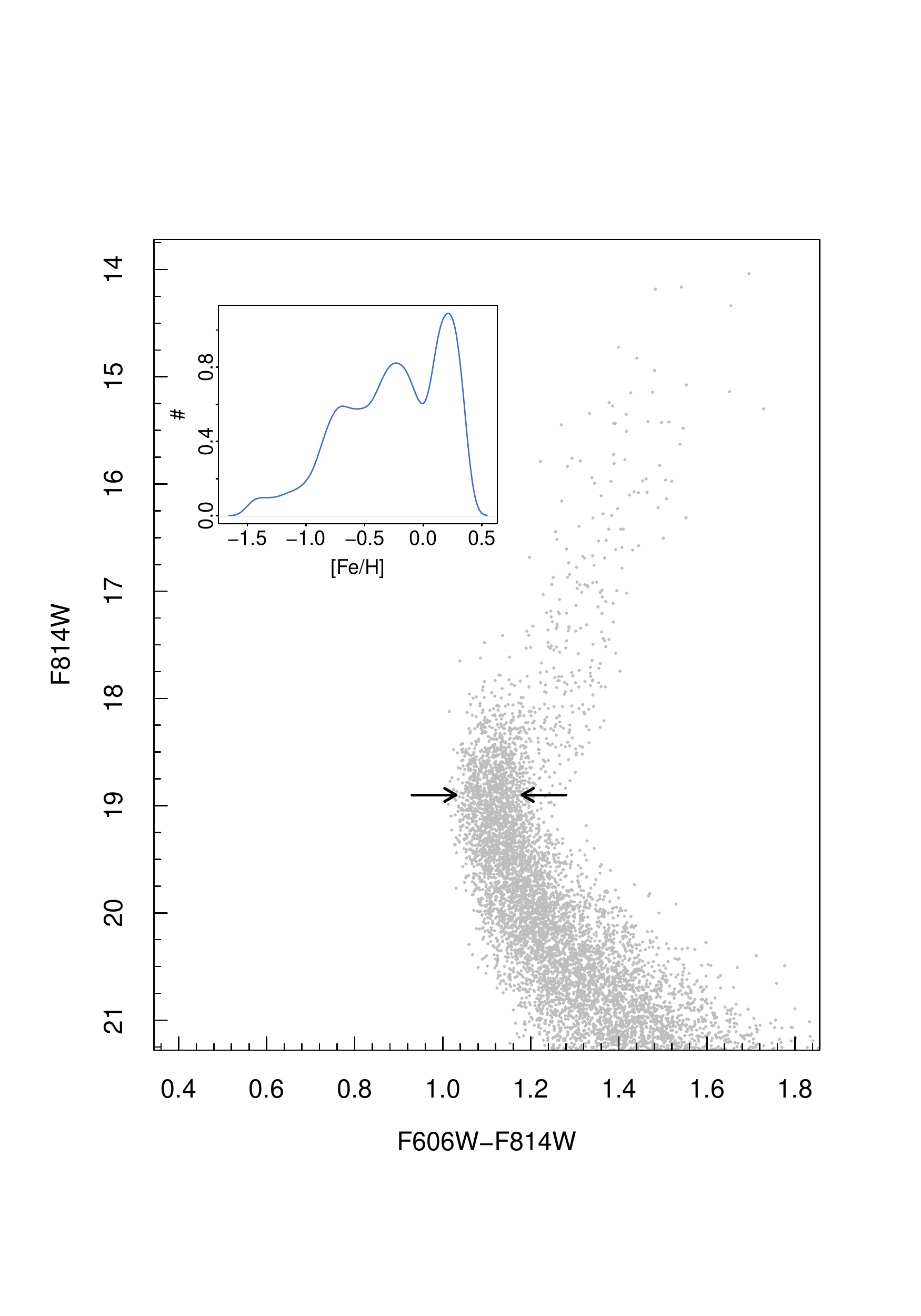}
\caption{Monte Carlo simulations  of the CMD
of Fig. \ref{fig:dhrmet} in the F814W/F606W-F814W bands, assuming the same conditions as in Fig. \ref{fig:hstdhr_amr}. 
The two arrows on each plot correspond to separation of 0.15. 
The inset frame shows
the corresponding metallicity distribution of the simulation.
}
\label{fig:hstdhr2}
\end{figure}

\begin{figure}
\includegraphics[trim=0 60 10 110,clip,width=9cm]{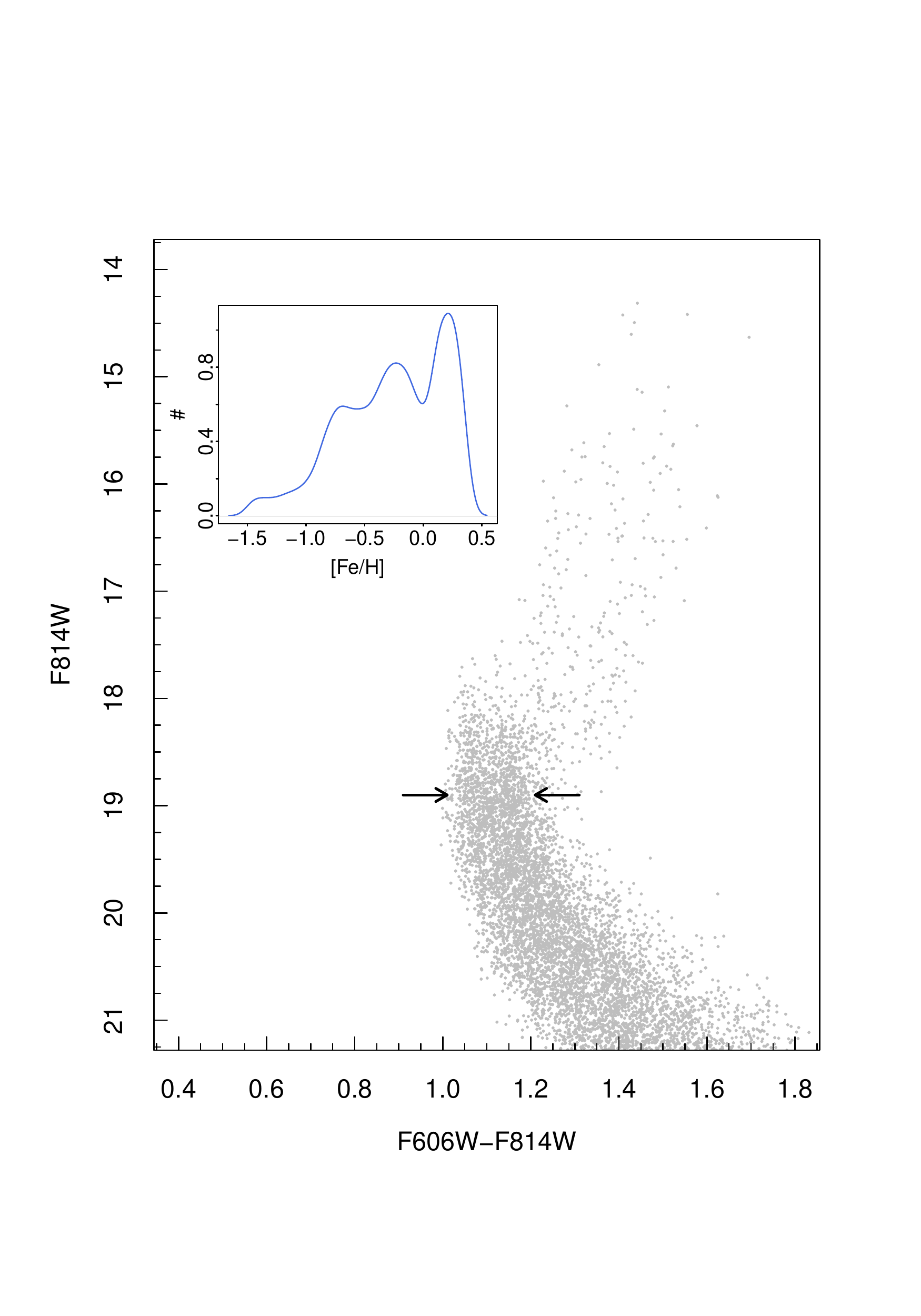}
\caption{Monte Carlo simulations  of the CMD Fig. \ref{fig:dhriso} in the F814W/F606W-F814W bands, assuming 
the same conditions as in Fig. \ref{fig:hstdhr_amr}. 
The two arrows on each plot correspond to separation of 0.20 mag. 
The inset frame shows
the corresponding metallicity distribution of the simulation.
}
\label{fig:hstdhr_iso}
\end{figure}

\section{The color-magnitude diagram of the Bulge}

We have built CMDs considering three different cases. In case (a) the AMR is represented by our model. 
In case (b), the bulge is considered exclusively old ($>$10 Gyr), but a relation still links 
age to metallicity, while in case (c), we considered that stars are born in a burst and all have an age of
11 Gyr.

\subsection{Isochrones}

Fig. \ref{fig:dhramr} (case a) shows a set of isochrones in the (F814W, F606W-F814W) plane, covering the age 
range from 2 to 13.5 Gyr. Their metallicity is chosen according to the AMR from our model
and represented in the inset frame.
The isochrones are calculated at the metallicity and alpha abundance content shown on each plot 
from the Yale library \citep{dem04}. The conversion to the ACS-WFC F814W, F606W bands 
uses colors from the Dartmouth database \citep{dot08}, which are based on the work of \cite{bed05}.
On Fig. \ref{fig:dhramr}, although the age spanned by the isochrones is 11.5 Gyr (2 to 13.5 Gyr), the range in color 
at the turn-off of the old populations (F606W-F814W) is no more than $\sim$0.08 magnitude between 
the isochrones at [Fe/H]=-1.5 (13.5 Gyr) and [Fe/H]=+0.35 dex (2~Gyr), respectively the bluest and reddest turn-offs in Fig. \ref{fig:dhramr}. 
Note that, because in case (a) the maximum
metallicity considered is +0.35 dex, we also limited the spread in metallicity in case (b) and (c) to the same value. 
No metallicities higher than 0.35 dex have been taken into account when generating the CMDs below. 
Although it has no effect on case (a) because stars more metal-rich than +0.35 dex are also young, so they do 
not increase the spread at the turn-off (see the 2 Gyr isochrone on Fig. \ref{fig:dhramr}), this selection underestimates the intrinsic
spread of case (b) and (c) -- by at most a few 0.01 mag -- strengthening the conclusions of this study.

Fig. \ref{fig:dhrmet} (case b) shows an intermediate case where the bulge is represented by an AMR 
that goes from [Fe/H]=-1.5 dex at 13.5 to [Fe/H]=+0.5 dex at 10 Gyr.  
It is possible that while being old ($>$10 Gyr), the bulge would have such an AMR. 
In this case, the range in color is about two
times the previous one, or about 0.17 mag. 

Fig. \ref{fig:dhriso} (case c) shows a set of 11 Gyr isochrones with different metallicities (from -1.5 to +0.5). 
The plot illustrates that a coeval population, having the range of metallicities similar
to what is observed for most of the bulge stars, has a wider turn-off ($\sim$0.2 mag) than the one presented
in Fig. \ref{fig:dhramr}, as expected.
Already from these simple comparisons, it is difficult to figure out how the case for an old, 
(both with an AMR or coeval) population can be compatible with observations, 
given that the observed photometric spread is smaller than 0.1 mag at the turn-off, 
while for case (b) and (c) it is larger than 0.17 mag.

Although a conclusion would be premature, because several other effects have to be taken into account 
before comparing with the observed CMD (see following section), these results do suggest that given the 
broad range of metallicities found in the bulge, a uniformly old population ($>$10 Gyr) would produce a 
turn-off wider than observed in the studies of CL08, \cite{cla11} and CA15.

Fig. \ref{fig:cmd_age} shows the set of isochrones of case (a) superposed on the CMD data of the SWEEPS
field\footnote{Based on observations made with the NASA/ESA {\it
Hubble Space Telescope}, obtained by the Space Telescope
Science Institute. STScI is operated by the Association of Universities for
Research in Astronomy, Inc., under NASA contract NAS~5-26555.}. 
Isochrones of age 13.5 and 2 Gyr are at the limit of the observed data points but we do not expect
much objects at these ages. Between 5 and 13 Gyr, the isochrones are well within the spread 
observed, and illustrate that case (a) is perfectly possible.

 Fig. \ref{fig:dartiso} shows different color plots from other isochrones \citep[Dartmouth database,][]{dot08}
both to confirm the expected spread in color given the metallicity spread observed in the bulge
using a set of different isochrones, but also to illustrate the expected 
intrinsic color range at the turn-off using different photometric bands from the HST-WFC3. 
The combination of F606W and F814W filters is most adapted to study this problem, because this color maximises 
the difference between the two extreme cases considered here (the case of a uniformly old population of age 11 Gyr
and that of a spread of $\sim$10 Gyr in age).
Hence, although in absolute values the spread at the turn-off is larger in the F390W-F814W filters (1.29 mag compared
to 0.21 mag in  F606W-F814W for the no-AMR case), the spread is only two times smaller in these filters
for the AMR case (from 1.29 to 0.59 mag), when it is three times smaller in F606W-F814W for the AMR case
(from 0.21 to 0.074 mag). So the contrast between the cases with and without AMR 
is expected to be the highest in the F606W-F814W bands.

\subsection{Simulated Color-magnitude diagrams and comparison with the observations}

We now illustrate this idea further by computing more realistic CMDs for the three cases presented in the previous section, 
taking into account the depth effect, the differential extinction and binaries through Monte Carlo simulations.
Starting from our model distributions, we simulate 
a set of (age, [Fe/H], [$\alpha$/Fe]) parameters through Monte-Carlo extraction from which we generate isochrones. 
A position along the isochrone is chosen according to the mass distribution 
generated from the IMF. 

An extraction of $\sim$ 10000 stars is made for each case considered. 
The 'intrinsic' dispersion (due to age and metallicity effects only, no dispersion due to depth effects, binaries,
differential reddening) of each simulation is 0.026 (case a), 0.036 (b) and 0.044 mag (c). Note that these dispersions
decrease to 0.023, 0.032 and 0.041 mag if only metallicities above -1.0 dex are selected, so the tail at low
metallicities has a limited effect. 
From these numbers, it is possible to see that the dispersion obtained for an isochrone 
population (case c) is hard to reconcile with the observations, being already $\sim$50\% larger than the observed
dispersion ($\sim$0.031 mag), given that none of the other sources of dispersion have been 
taken into account.

In order to obtain more realistic estimates, various contributions 
must be taken into account. The first one is the depth effect:
all stars are not at the same distance, and a variation of $\pm$1 kpc
at 8kpc will translate into a difference in the distance modulus of about
(+0.26,-0.29) mag. 
Following Valenti et al. (2013), a dispersion of 0.26 mag 
has been added to take into account the spread
in distance of stars in the bulge. 
The dispersion now rises to 0.035, 0.042 and 0.051$\pm$ 0.001 mag, or a spread of 0.105, 0.126 and 0.153 mag. 

Another potentially important effect that one has to take into account comes from 
differential reddening. In CA15, this is taken into account by assuming a 
possible reddening variation of $\pm$0.05 mag over the whole field,
which we model by assuming a dispersion in reddening of 0.0166 mag.
The resulting new dispersions are now 0.039$\pm$0.0013, 0.045$\pm$0.0009 and 0.054$\pm$0.0012 mag, or a spread of 0.117, 0.135 and 
0.162 mag for case (a), (b) and (c) respectively (if metallicities are limited to [Fe/H]$>$-1.0 dex, the dispersions are
0.037$\pm$0.0010, 0.043$\pm$0.0011 and 0.051$\pm$0.0012mag).

The percentage of binaries and the distribution of their mass ratios is not known for the bulge. 
We simulate binaries assuming a uniform distribution of mass ratios \citep{rag10}. Because the Yale 
stellar track library is limited to stellar masses higher than 0.4 M$_{\odot}$, we derived unresolved
binary parameters only in cases where the companion has a mass higher than this limit.
Because the mass at the turn-off is about 0.8 M$_{\odot}$, it implies that companions with 
an absolute magnitude difference higher than about 4.5 mag are not modelled, corresponding to neglecting 
effects on combined magnitudes that are smaller than 0.015 mag, which is reasonable. 
According to CA15, the minimum fraction of binaries is about 30\%, but it would easily reach 50\%
if it was similar to the disk (and if indeed the bulge is essentially made of disk stars).
Assuming 30\% of binaries, the dispersion increases to 0.051$\pm$0.001 mag (case a), 0.056$\pm$0.0019 
(case b) and  0.064$\pm$0.0015 mag (case c). 
These numbers rise to 0.056$\pm$0.0014, 0.058$\pm$0.0017 mag and 0.065$\pm$0.0014 in the case of 50\% binaries, 
and 0.056$\pm$0.0013 mag (case a) 0.059$\pm$ 0.0014 (case b) and 0.066$\pm$0.0015 mag in the case of 100\% of binaries. 
As can be seen, the increase in dispersion is not linear with the fraction of binaries, but varies most
rapidly with small percentages, then flattens above 50\%. 

We conclude that even in case (a) the dispersion is higher than the one measured on the SWEEPS data, which means we have 
probably overestimated one or several of the sources of dispersion. 
Even so, it is difficult to imagine that the dispersion in cases (b) and (c) can be reconciled with a dispersion 
lower than 0.03 mag as the one observed, being already larger than this limit when only age and metallicity are 
taken into account. 
The freedom that we have to accommodate for a smaller dispersion in case (a) is larger because, in addition to 
the possible overestimated differential reddening, distance uncertainty and effect of binaries, we have not tuned 
the AMR in any way, nor the SFH, which both play a role in determining the dispersion at the turn-off.

In addition, we also note that the spread in color of the giant branch is also remarkably tight in 
Clarkson et al. (2011, their Fig. 1), being limited to at most 0.2 mag. This is also in much better 
agreement with case (a) than with case (c), where it can be seen that the spread is about 0.4 mag. 

Hence, the present study proves that the bulge must contain young disk stars 
in substantial amount if we want the spread of metallicities observed in several independent 
studies to be compatible with an apparently uniformly old bulge as observed from its CMD.
If the bulge contains metal-rich stars, these have to be significantly younger than 
the subsolar metallicity population.

A final comment is worth making. 
We find that the CMD is not compatible with a unique (old) age given the spread in the metallicities, 
but this statement is true at all ages. In other words, the observed CMD is also a constraint on the 
dispersion of the age-metallicity relation in the inner-disk and bulge, and tells us that this relation 
must be relatively tight in order to accommodate the spread in color at the turn-off in the SWEEPS CMD.
This is at variance with the age-metallicity relation observed at the solar vicinity for example but 
in accordance with the closed-box model we assumed for the inner disk and bulge.
It tells us that the age-metallicity relation in the inner disk must be tighter, contrary to what is seen in the
solar vicinity. Our interpretation is that, the solar vicinity being in the transition region between the inner and
outer disk, the metallicity gradient is very steep locally, bringing stars of significantly different metallicities at
a given age in the local volume, while in the inner disk the evolution has been much more homogeneous.

\section{Discussion}

We have shown that:

(1) A coeval stellar population with an age of 11 Gyr and a metallicity
distribution compatible with that of the Baade's Window implies a spread in the turn-off color
significantly larger than the one observed in the SWEEPS field.

(2) Even for a bulge with an age older than 10 Gyr and in which metallicity correlates 
with age, the turn-off region should be significantly broader than what is observed in the SWEEPS field.

(3) A bulge population which is assumed to follow the same chemical characteristics 
(and in particular the same AMR)
as those we used to model the inner disk (which is expected if the bulge is a bar formed from the disk) 
produces a CMD with a tight turn-off, hiding a spread in age of about 10 Gyr, between 3 and 13 Gyr.

These results imply that the presently known constraints from CL08 and \cite{val13}
are not evidence that the bulge is exclusively old ($>$10 Gyr). On the contrary, they are evidence 
that the more metal-rich stars in the bulge must be significantly younger than the metal-poor ones, 
otherwise we would expect the turn-off of the bulge to be significantly broader than observed.

It is worth emphasizing here that we still view the bulge as predominantly old:
in our model (see Haywood et al. 2016, in preparation), about 50\% of the stars must have ages greater than
9 Gyr. In the CMD of Fig \ref{fig:hstdhr_amr}, the percentage decreases to 34\%, because we considered the SLEEPS field
would have a negligeable amount of stars younger than 3 Gyr. 
Clearly, there is room for a younger population, which makes the bulk of the metal-rich peak. 
This population originate in the inner thin disk \citep[see][for a review]{dim16},
and its stars formed after the quenching episode \citep{hay16} that occurred at the end of 
thick disk phase, and so must be essentially younger than 8 Gyr. 

Before discussing the evidence from the literature in favor of an exclusively old bulge, it is worth
discussing the ages of the metal-rich dwarfs from B13 in the light 
of our results, and to see in particular how far the age of their metal-rich stars are 
still compatible, within the error bars, with being old ($>$ 10 Gyr), supposing they are all bulge stars. 
In their Fig. 4, 
out of 19 stars having [Fe/H]$>$0.1 dex, 14 have ages compatible with being 5 Gyr old
or younger within 1 sigma on LogT$_{eff}$ or log~g, while only 5 are within one sigma of being 
10 Gyr or older. All these 5 stars are also compatible with being younger than 10 Gyr (and for 
3 of them, younger than 5 Gyr) within 1 or 2 sigmas. 
On the contrary, 9 stars younger than 10 Gyr are irreconcilably hotter than the 10 Gyr isochrone,
being sometimes off by more than 500 K (4 to 5 $\sigma$ on temperature). 
This is to emphasize the fact that, while all metal-rich stars of B13 are 
reasonably (within 1 or 2 sigmas) compatible with being young (5 Gyr or younger), the reverse is not
true, and no young star in B13 is compatible with being 10 Gyr old or older (within less than 
4 to 5 sigmas on $\rm LogT_{eff}$), although this would be expected if the bulge was exclusively old.
We now review the arguments that have been invoked to support that the metal-rich 
population must be exclusively old.

(1) The stellar population arguments

Up to the nineties, the Galactic bulge has been essentially associated to an old, 
collapsed component \citep[e.g.][]{ric90,mat90} and it has been suggested that even its metal-rich 
stars could be older than the halo component \citep{ren90}.
Two results mainly have concurred to support the idea
of a uniformly old bulge, even though the possibility of young super metal-rich stars has been recognized early
\citep{woo83}. The first is the prediction by Matteucci \& Brocato (1990) that alpha-enhanced
abundances were to be expected even at high metallicities, and the first observations supported 
this prediction \citep{mcw94}. The second is the idea that the angular momentum distribution of the bulge
is similar to the one of the halo -- the thick disk having similar distribution to the thin disk -- 
thereby linking the bulge to a primordial collapse of the Galaxy, see e.g \cite{wys92}.
While we discuss the chemical argument in more detail below, 
it is worth emphasizing that subsequent studies have not confirmed the substantially enhanced 
alpha abundances at solar metallicities and above. 

The discovery that the bulge is in fact a bar \citep{bli91}, and the discovery
that vertical instabilities create boxy and peanut shape morphologies \citep{com81} has prompted 
the idea that at least part of the 
bulge could be made of disk stars trapped into the bar, with the dual nature of the bulge 
already suggested at this time \citep{ric93}. 
However, the measurement of alpha-enhanced
abundances at high metallicities \citep{mcw94, lec07} and the seemingly old CMD (CL08) 
suggested up to recently that the bulge was a mixture of a classical and pseudo-bulge, prompting 
the presentation of more sophisticated chemical evolution models with uniformly old bulge, even 
at high metallicities \citep{bal07,mcw08}.

It has been argued at the same time from kinematic arguments that if a classical bulge exist in the MW, it must either
be small, or inexistent (\citealp{she10}, \citealp{kun12}, \citealp{dim14}, \citealp{kun16}  estimated
its mass to be 1\% of the total mass of the inner Galaxy). 
If the classical bulge is small, it implies that the bulge of the MW must be a pseudo-bulge, 
that is a bulge formed from the disk through dynamical instabilities. 
The evidence that most of the bulge is a pseudo-bulge is now overwhelming. 
The last evidence came from the finding by \cite{mcw10} and \cite{nat10} that the 
bulge is X-shaped. 
It has been shown in \cite{dim14,dim15}  how the disk inside the Outer Linblad resonance radius 
can be mapped into the bar and how the stellar components detected by \cite{nes13}
can be identified to the thick and thin disks of the MW \citep{dim16},
while the peanut-shape is seen to be most prominent in the metal-rich component \citep{nes13}, clearly
pointing to a disk origin. 

This rapid summary shows that within a little more than a decade, new data on the 
structural shape and kinematical characteristics have changed our view of the bulge, from 
a collapsed, rapidly formed spheroidal component to a bar formed from dynamical instabilities in the 
disk. 
These arguments are not direct evidence that the metal-rich objects must be younger than 
ages expected for an old spheroid formed by mergers (as often advocated for a classical bulge), but 
they are evidence of a disk origin of these stars. Because there is no clue that metal-rich 
stars in the disk must be older than 10 Gyr, it is reasonable to expect those seen in the 
bulge to be also younger than this limit.

(2) The chemical argument

As mentioned previously, it has been argued that, because the 
bulge is $\alpha$-rich compared to the solar vicinity, it formed on a short time scale \citep[e.g.][]{zoc06, gri12}.
While it has been found in the first studies that metal-rich bulge stars seemed to be systematically
more alpha-enhanced than local disk stars, the offset has only been decreasing in the last ten years. 
Most recently, studies made on the same type of stars (dwarfs) by the same group using the same method (B13 and Bensby et al. 2014) obtained results that are almost indistinguishable. 

\begin{figure}
\includegraphics[width=9.cm]{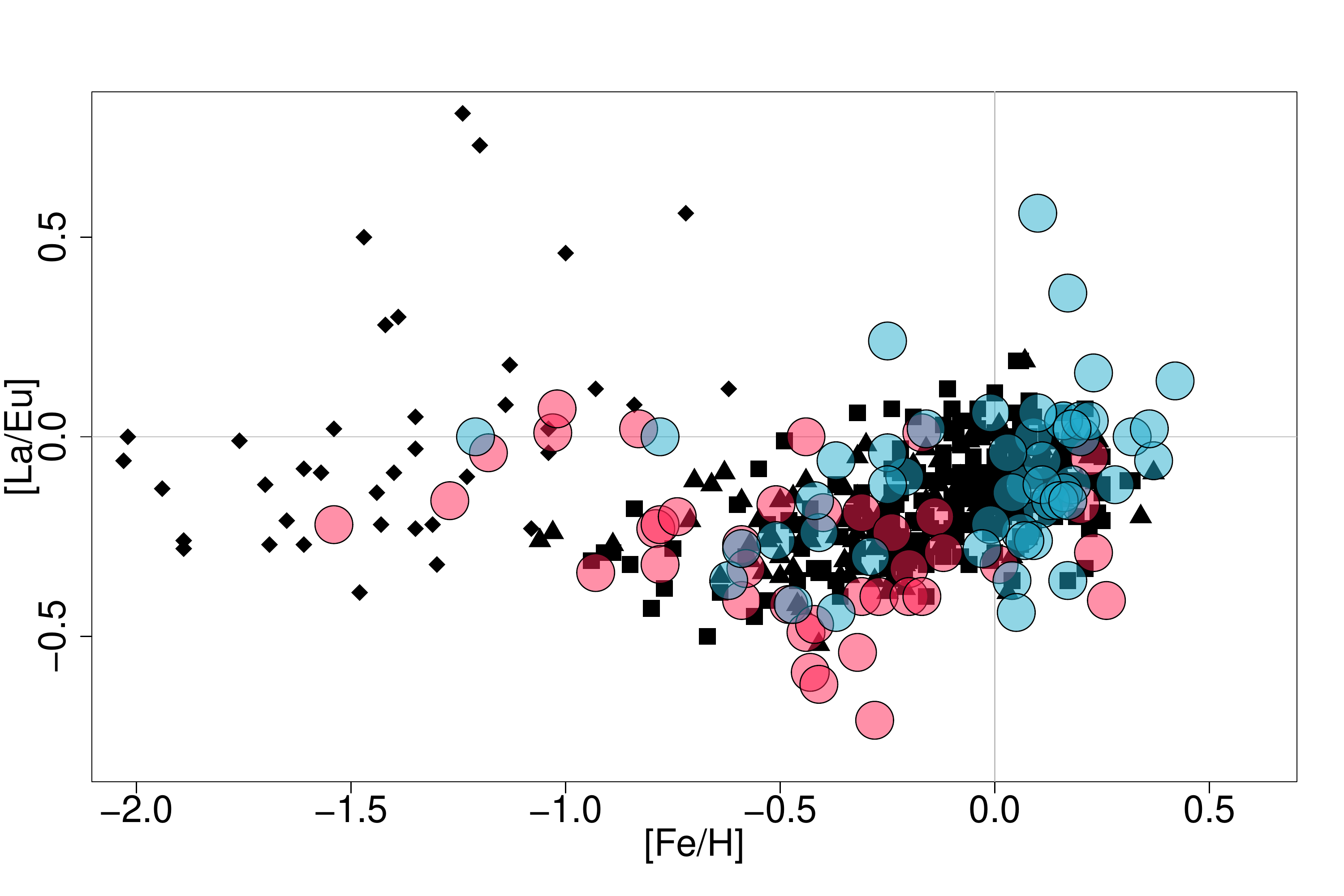}
\caption{[La/Eu] from \cite{joh12} and \cite{van16} (respectively pink and 
blue larger symbols) of bulge stars compared to data from \cite{bat16} (triangles), \cite{mis13}
(squares) and \cite{ish13} (diamons) for solar vicinity stars. 
There is no evidence from this data that the bulge and the solar vicinity  distributions are different.
}
\label{fig:eula}
\end{figure}

\cite{joh12} argued that the low ratio of [La/Eu] indicates that the majority of the bulge formed rapidly
($<$1Gyr). Fig. \ref{fig:eula} compares the  [La/Eu] abundance ratio for two different bulge samples from \cite{joh12} and 
\cite{van16}, together with three different samples of solar vicinity stars from \cite{bat16},
\cite{mis13} and \cite{ish13}. They all show good agreement, and the two bulge
samples show no evidence of having lower [La/Eu] than solar vicinity stars. 
We conclude that the available [La/Eu] abundance ratios do not provide evidence that the bulge formed 
faster than the inner disk. Other arguments are put in the balance for an old bulge by \cite{nat15}, 
such as the presence of RR Lyrae and old globular clusters. However, these only confirm
(if it was needed) that there are old  populations in the bulge, but does not imply 
that the bulge must be exclusively old.

More generally, it is important to emphasize that there is no argument coming from chemical evolution modelling
asserting that the metal-rich population must be old. 
The original results from \cite{mat90} are based on the endorsing a uniformly old age for the bulge
(and particularly for metal-rich stars), reflecting a generally accepted idea
of that time\footnote{This subject was then pretty much in the air, see for example \cite{fro90}: 'Whether or not
the Galactic bulge contains a detectable population of stars with age much less than 10 Gyr is a subject of considerable debate', 
to conclude that the bulge has no evidence for a population younger than 10 Gyr.}.
The most recent Galactic Chemical Evolution (GCE) models are still trying to cope with representing the bulge as exclusively old, and the HST
CMD of a seemingly old population has a heavy weight in this choice. 
Although they reproduce reasonably well the observed chemical constraints,  
we want to emphasize that: (1) these models see the bulge as a component isolated from the disk, 
which is unrealistic given the present evidence of the bulge being mostly a pseudo-bulge,
(2) Modelling the bulge as an exclusively old and isolated system imposes a complexity and a number
of adhoc parameters which are actually not required by the data if the bulge is considered as 
a bar grown from the disk and having the age range of the disk. 
In \cite{gri12}, the metal-rich component is modelled with parameters different than those used for 
stars below [Fe/H]$<$0.0 dex. 
The infall time scale, star formation efficiency, and IMF are assumed to be different than that 
of the stars at lower metallicities.
In \cite{tsu12}, the metal-rich component is formed from gas at a metallicity [Fe/H]$\sim$-0.3 dex,
which, having to form stars on a short time scale, also implies a top-heavy IMF. 
GCE models are able to produce the metal-rich component as being exclusively old, 
but this comes with a price, because
forming stars enriched to [Fe/H]$>$+0.3 dex in a limited amount of time (within a few Gyr) is challenging.
This is usually done by endorsing a flatter IMF than the one used for the solar vicinity, for instance a
Salpeter IMF, or even flatter \citep{mat99, bal07, tsu12, gri12}.
In this context, it is worth mentioning that CA15 measure an IMF for the bulge
population which is compatible with that of the disk, and different from the Salpeter IMF. 
Also, while it is sometimes recognised in these studies that the (metal-rich) bulge is 
a bar formed from disk material, these models still represent this component with specific
and adhoc parameters which differ from those adopted to reproduce the metal-poor component, 
but with little justification. For example, in \cite{gri12}  metal-rich stars are formed within less than 1 Gyr after the 
metal-poor population, but no argument justifies that the star formation efficiency is decreased by 
more than a factor of 10 (from 25 Gyr$^{-1}$ to 2 Gyr$^{-1}$) in such short amount of time, or
that the time scale for infall is increased 
by a factor of 30 (from 0.1 to 3 Gyr).

\section{Conclusions}

We conclude this study by saying that there is no evidence from the CMD in the SWEEPS field that the bulge is exclusively old. 
The HST CMD has been claimed to provide the best evidence of an old bulge, but it is in fact evidence of the contrary:
the tight turn-off of the CMD is possible only if it is hiding a correlation between age and metallicity, 
which means that metal-rich stars ([Fe/H]$>$0) must be significantly younger than the metal-poor population ([Fe/H]$<$0. dex).
In Haywood et al. (in preparation), we show that a model which represents the bulge with a thick 
and a thin disk reproduces well observed chemical constraints, with no additional parameters than those used
to represent the thin and thick inner disks.
Evidently, since we model the bulge with a thick+thin disk populations, we release the constraint of having ages 
greater than $\sim$ 10 Gyr only. This makes the production of metal-rich stars much easier, and in particular
there is no need to assume an IMF different from that of the disk.
In our model, stars having [Fe/H]$>$0 dex are all younger than 8 Gyr, and are formed after the quenching 
episode (Haywood et al. 2016). 
Our results show that unless the MDF in the SWEEPS field is strongly different from that 
of the neighbouring Baade's Window, these younger stars are necessary to reproduce the 
CMD obtained from the HST.

\begin{acknowledgements}
We are grateful to the referee for a detailed report and very helpful comments.
A.C. acknowledges NASA which supported her work through grants GO-9750 and GO-12586 from the Space
Telescope Science Institute, which is operated by AURA, Inc., under NASA contract NAS~5-26555.
\end{acknowledgements}

\clearpage

\end{document}